\documentstyle[12pt]{article}
\begin{document}
\thispagestyle{empty}
\begin{flushright}
INR 0973/98 \\
February 1998
\end{flushright}
\bigskip\bigskip\bigskip
\vskip 2.5truecm
\begin{center}
{\LARGE {Extension of the Nielsen-Ninomiya theorem}}
\end{center}
\vskip 1.0truecm
\centerline{S.~V.~Zenkin}
\vskip5mm
\centerline{Institute for Nuclear Research of
the Russian Academy of Sciences}
\centerline{60th October Anniversary Prospect 7a, 117312 Moscow,
Russia}
\centerline{E-mail: zenkin@al20.inr.troitsk.ru}
\vskip 2cm
\bigskip \nopagebreak \begin{abstract}
\noindent
The index theorem is employed to extend the no-go theorem for lattice
chiral Dirac fermions to translation non-invariant and non-local
formulations.\\


\end{abstract}
\vskip 1.5cm

\newpage\setcounter{page}1

{\bf 1.} Lattice regularization of functional integrals is the basis
of the most powerful methods of non-perturbative treatment of
non-abelian gauge theories. As any regularization, the lattice breaks
some properties of the underlying continuous theory. The
Nielsen-Ninomiya theorem \cite{NiN} imposes non-trivial limitations
on those properties of the fermion action that can be maintained on
the lattice. In particular, this theorem states that if a lattice
Dirac operator provides correct free fermion spectrum in the
continuum limit, it cannot simultaneously be (i) lattice translation
invariant, (ii) local, and (iii) chirally invariant (for a more
detailed formulation, see \cite{Lu}).

Among the properties (i)--(iii) chiral symmetry is of particular
interest due to the role it plays both in the low energy dynamics of
QCD and in the construction of chiral gauge theories. Various
attempts have been made to hold chiral invariance on the lattice at
the cost of losing translation invariance \cite{tni}, locality
\cite{nl}, or the correct free fermion spectrum \cite{Za}.

At the same time, one can hardly consider a lattice definition of
Dirac operator as satisfactory if it does not reproduce, exactly or
in some limit, all known exact properties of the continuum Dirac
operator. One of such non-trivial properties is a consequence of the
Atiyah-Singer index theorem \cite{AtS}. This is a relation between
the number of chiral zero modes of the Dirac operator in a smooth
external gauge field on a compact manifold and the topological number
$Q$ of this external field, so that on the manifolds of index zero
(like a sphere and a torus) one has \cite{Sc}
\begin{equation}
Q = n_- - n_+,
\label{eq:it}
\end{equation}
where $n_{+}$ ($n_{-}$) is the number of zero modes of positive
(negative) chirality. Note that on a finite torus relation
(\ref{eq:it}) is non-trivial also in the abelian case (see, for
example, \cite{SmV}).

Inability of some particular lattice Dirac operator to reproduce this
property were discussed in \cite{zi}. The crucial role of the index 
theorem in defining both vector and chiral gauge theories on a lattice 
was emphasized in \cite{NaN}.

The aim of this paper is to demonstrate that by requiring relation
(\ref{eq:it}) to hold, at least approximately, on a finite lattice 
one can extend the no-go theorem to generic chirally invariant lattice 
Dirac operators including translation non-invariant and non-local ones, 
provided they satisfy a mild spectral condition. \\

{\bf 2.} We consider a gauge theory defined on a {\em finite lattice}
with the fermion action
\begin{equation}
S_f = \sum_{m, n} \overline{\psi}_m D_{m n}(A) \psi_n,
\label{eq:A}
\end{equation}
where $A$ stands for gauge variables. The lattice may be non-regular
with arbitrary boundary conditions. The Dirac operator $D$ may be
translation non-invariant or non-local, it also may not be gauge
invariant.

The generic form of the Dirac operator in the chiral representation
of $\gamma$-matrices, $\gamma_5 = \mbox{diag}(I, -I)$,
$\gamma_{\mu}^{\dagger} = \gamma_{\mu}$, reads as
\begin{equation}
D = \pmatrix {M_{+} & D_{-} \cr
D_{+} & M_{-} \cr}.
\label{eq:D}
\end{equation}
Matrices $D_{+}$ and $D_{-}$ are lattice transcriptions of the
covariant Weyl operators $\sigma_{\mu} (\partial_{\mu} + i A_{\mu})$
and $\sigma_{\mu}^{\dagger} (\partial_{\mu} + i A_{\mu})$,
respectively, where $A_{\mu}$ is the gauge field, $\sigma_{\mu} = (1,
i)$ in two dimensions and $\sigma_{\mu} = (I, i\sigma_i)$ in four
dimensions. Matrix $M = \mbox{diag}(M_{+}, M_{-}) \neq \mbox{constant} 
\times I$ is the measure of breaking of chiral symmetry: 
$D \gamma_5 + \gamma_5 D = 2 M \gamma_5$.

We now assume that the chirally invariant part of $D$ is normal, i.e.
$(D - M)(D - M)^\dagger = (D - M)^\dagger (D - M)$. This implies that
\begin{eqnarray}
&D_{+}^{\dagger} D_{+} = D_{-} D_{-}^{\dagger}, \cr
&D_{+} D_{+}^{\dagger} = D_{-}^{\dagger} D_{-},
\label{eq:no}
\end{eqnarray}
and that operator $D - M$ has a complete orthonormal set of
eigenvectors.  This is {\em the only condition} we impose on the form
of the Dirac operator. Note that (\ref{eq:no}) is automatically
satisfied, if $D_{\pm}$ obey the same relation as the corresponding
continuum operators: $D_{-} = -D_{+}^{\dagger}$.

Now everything is fixed to prove the following

\noindent {\bf Theorem.} {\em A necessary condition for the lattice
Dirac operator (\ref{eq:D}), (\ref{eq:no}) to reproduce a nonzero
index (\ref{eq:it}) is $M \neq 0$.}

We give two proofs: a `mathematical', that uses only properties of
linear operators in finite-dimensional Hilbert spaces, and
`physical', that uses customary technique of quantum field theory.
Both are very simple.

{\em Proof 1 (`mathematical').} If $M = 0$, by definition $n_{+} -
n_{-} = \mbox{dim ker }D_{+} - \mbox{dim ker }D_{-}$, where $\mbox{dim
ker }D_{\pm}$ mean the dimensions of the kernels of the operators
$D_{\pm}$. By virtue of condition (\ref{eq:no}) and finiteness of the
lattice, one has: $\mbox{dim ker }D_{+} = \mbox{dim
ker }D_{+}^{\dagger} D_{+} = \mbox{dim ker }D_{-} D_{-}^{\dagger} =
\mbox{dim ker }D_{-}^{\dagger} = \mbox{dim ker }D_{-}$.  Hence, $n_{+}
- n_{-} = 0$.

{\em Proof 2 (`physical').} Define the partition function in an
external gauge field
\begin{equation}
Z_f = \int \prod_{n} d\psi_n d\overline{\psi}_n \exp(-S_f
- \epsilon \sum_n \overline{\psi}_n \psi_n)
= \mbox{det} (D + \epsilon),
\label{eq:Z}
\end{equation}
and the fermion expectation values
\begin{equation}
\langle O(\psi, \overline{\psi}) \rangle_f
= Z_{f}^{-1} \int \prod_{n} d\psi_n d\overline{\psi}_n
O(\psi, \overline{\psi}) \exp(-S_f
- \epsilon \sum_n \overline{\psi}_n \psi_n),
\label{eq:ev}
\end{equation}
where $\epsilon$ is an infinitesimal mass introduced to avoid possible
singularities.

The Ward identity associated with the singlet chiral transformations
$\psi_n = \exp(i \alpha_n \gamma_5) \psi'_n$, $\overline{\psi}_n =
\overline{\psi'}_n \exp(i\alpha_n \gamma_5)$ \cite{KSS} reads as
\begin{equation}
\left. i \frac{\partial}{\partial \alpha_n} \ln Z_f \right|_{\alpha =
0 } = \langle \overline{\psi}_n
\sum_m \gamma_5 D_{n m} \psi_m + \sum_m \overline{\psi}_m
D_{m n} \gamma_5 \psi_n \rangle_f + 2 \epsilon \langle
\overline{\psi}_n \gamma_5 \psi_n \rangle_f = 0.
\label{eq:wi}
\end{equation}
Taking the sum over $n$ we obtain
\begin{equation}
\sum_{m n} \langle \overline{\psi}_m M_{m n} \gamma_5 \psi_n
\rangle_f + \epsilon \sum_n \langle \overline{\psi}_n
\gamma_5 \psi_n \rangle_f = 0.
\label{eq:swi}
\end{equation}

Now let $M = 0$. Then, by our condition $D$ is normal and
anticommutes with $\gamma_5$. Therefore it has complete orthonormal
set of eigenfunctions such that if $f_\lambda$ is an eigenfunction
with the eigenvalue $\lambda$, the function $\gamma_5 f_\lambda$ is
an eigenfunction with the eigenvalue $-\lambda$, and zero modes $f_0$
can be chosen to be chiral, i.e. such that $\gamma_5 f_0 = \pm f_0$.
Then, one has
\begin{equation}
\epsilon \sum_n \langle \overline{\psi}_n \gamma_5 \psi_n \rangle_f =
- \epsilon \sum_{f_{\lambda}, n} \frac{f^{\dagger}_{\lambda}(n) \gamma_5
f_{\lambda}(n)}{\lambda + \epsilon} = n_{-} - n_{+} = 0,
\label{eq:zi}
\end{equation}
independently of the configuration of the external gauge fields.

\noindent{\em QED}.\\

{\bf 3.} Let us make some comments:

If on a finite lattice one has $n_{+} - n_{-} = 0$, then a non-zero
result cannot be obtained in any limit. If $M \neq 0$, the l.h.s. in
(\ref{eq:zi}) generally does not equal $n_{+} - n_{-}$ and may not be
integer. However, in such a case relation (\ref{eq:it}) can be
approached in some limit, as it happens in the case of Wilson
fermions \cite{SmV} (for more references see, for instance, 
\cite{GaH}).

Thus these considerations exclude all chirally invariant Dirac 
operators defined on a finite lattice and satisfying condition 
(\ref{eq:no}).

The index theorem (\ref{eq:it}) is closely related to the existence
of the anomaly in the divergence of the singlet axial current
\cite{ABJ}. On a lattice the current may be defined up to the terms
vanishing in the naive continuum limit. Therefore chiral invariance
of a lattice Dirac operator does not exclude a priori the appearance
of the anomaly at some definition of the current. By employing the
global relation (\ref{eq:it}) we completely avoid this ambiguity.

Although the considered theorem directly concerns only Dirac
fermions, it has some bearing on the construction of chiral gauge
theories. For instance, it demonstrates that at least one of the
operators $D_{\pm}$ satisfying condition (\ref{eq:no}) always fails
to reproduce all the properties of its continuum counterpart. In
particular, this means that to define a Weyl fermion on a lattice it
is not sufficient only to define an anti-hermitean lattice
transcription of the covariant derivative $\partial_{\mu} +
i A_{\mu}$ (see also \cite{NaN}).

The condition $M \neq 0$ does not necessarily lead to the
complications typical of Wilson fermions. Very interesting examples
are the operators constrained by the Ginsparg-Wilson condition
\cite{GiW}--\cite{Ne1}: $2 M \gamma_5 = r D \gamma_5 D$, where $r$ is
a non-zero parameter. In fact, this condition ensures a non-trivial
realization of exact chiral symmetry on a lattice \cite{Ha},
\cite{Lu}, at least in some region of phase space \cite{Ne2}. 
It is an interesting
question whether the properties of the Dirac operators satisfying the
Ginsparg-Wilson relation can facilitate the construction of chiral
gauge theories. Note that finite lattice versions of chirally invariant 
non-local fixed-point action \cite{GiW}, \cite{Wi} corresponding to 
$r = 0$ should be rejected. An example of the
Dirac operator that, despite $M \neq 0$, has gauge invariant modulus
of the determinant when the gauge coupling is chiral is proposed in
\cite{Ze} (where this operator has been incorrectly called chirally
invariant). In the free fermion limit this operator obeys the
condition $M (D - M) = 0$. \\

I am grateful to M.~L\"uscher for a clarifying comment on the
conditions of the Nielsen-Ninomiya theorem, and to T.~W.~Chiu,
H.~Neuberger and J.~F.~Wheater for informing me of earlier related 
discussions.

\end{document}